\documentclass[journal]{IEEEtran} 
\ifCLASSINFOpdf
\else
\fi
\usepackage{cite}
\usepackage{amsmath}
\usepackage{amssymb}
\usepackage{amsfonts}
\usepackage{psfrag}
\usepackage{epsfig} 
\usepackage{algorithm}
\usepackage{algorithmic}

\newcommand{\B}[1]{\boldsymbol{#1}}
\begin{document}
%
\title{Blind Estimation of Sparse Broadband Massive MIMO Channels with Ideal and One-bit ADCs} 

\author{\IEEEauthorblockN{Amine Mezghani,~\IEEEmembership{Member,~IEEE} and A. Lee Swindlehurst,~\IEEEmembership{Fellow,~IEEE}} \\
\thanks{A. Mezghani was with the University of California, Irvine, CA 92697 USA, and is now with the University of Texas at Austin, TX 78712 USA (e-mail: amine.mezghani@utexas.edu). A. L. Swindlehurst is with the Center for Pervasive Communications and Computing, University of California, Irvine, CA 92697 USA, and is also a Hans Fischer Senior Fellow of the Institute for Advanced Study at the Technical University of Munich (e-mail: swindle@uci.edu).}
\thanks{This work was supported by the National Science Foundation under
ECCS-1547155, and by the Technische Universit\"at M\"unchen Institute for
Advanced Study, funded by the German Excellence Initiative and the European
Union Seventh Framework Programme under Grant No. 291763, and
by the European Union under the Marie Curie COFUND Program.}
\thanks{Some parts of this work have been published in \cite{Mezghani_2017}, where the unquantized and flat fading scenario was considered.}
}


%


\maketitle

\begin{abstract}
We study the maximum likelihood problem for the blind estimation of massive mmWave MIMO channels while taking into account their underlying sparse structure, the temporal shifts across antennas in the broadband regime, and ultimately one-bit quantization at the receiver. The sparsity in the angular domain is exploited as a key property to enable the unambiguous blind separation between  user's channels. The main advantage of this approach is the fact that the overhead due to pilot sequences can be dramatically reduced especially when operating at low SNR per antenna. In addition, as sparsity is the only assumption made about the channel, the proposed method is robust with respect to the statistical properties of the channel and data and allows the channel estimation and the separation of interfering users from adjacent base stations to be performed in rapidly time-varying scenarios. For the case of one-bit receivers, a blind channel estimation is proposed that relies on the Expectation Maximization (EM) algorithm. Additionally, performance limits are derived based on the clairvoyant Cram\'er Rao lower bound.  Simulation results demonstrate that this maximum likelihood formulation yields superior estimation accuracy in the narrowband as well as the wideband regime with  reasonable computational complexity and limited model assumptions.

\end{abstract}


%
\IEEEpeerreviewmaketitle

\section{Introduction}  
Channel estimation is recognized as one of the key issues in developing the fifth generation of wireless communication systems \cite{eetimes}. In particular, estimating massive MIMO millimeter wave (mmWave) channels is challenging due to the larger dimensions, larger bandwidths, hardware imperfections and faster temporal variations. In addition, such systems are expected to operate at low SNR values per antenna due to several factors like increased path-loss, hardware restrictions of the power amplifiers, larger noise bandwidths and smaller antenna sizes, which, together with the issues of pilot-contamination and carrier frequency offset, renders common pilot based estimation methods inefficient and even impossible. \\

The high power consumption and complexity of massive MIMO systems have led researchers to consider receivers with low-resolution quantizers 
\cite{nossek,ivrlac,ivrlac2006,mezghani2007,mezghaniisit2007,mezghani_ICASSP2008,mezghaniisit2008,Mo_2016,Zymnis_2010,mezghani_itg_2010,Yongzhiuplink,mollen2016performance,juncil2015near,jacobsson2015one,hea14,ning2015mixed}.
In the extreme case, a one-bit A/D-converter (ADC) can be employed that consists of a simple comparator and consumes negligible power. One-bit ADCs do not require an automatic gain control and the gain stages needed prior to them are substantially reduced \cite{donnell}. Ultimately, one-bit conversion is, in view of the current CMOS technology, the only conceivable option for direct mmWave bandpass sampling close to the antenna, eliminating the need for power intensive RF components such as mixers, frequency synthesizers and local oscillator distribution networks  \cite{donnell}. In addition, the use of one-bit ADCs not only simplifies the interface to the antennas by relaxing the RF requirements but also simplifies the interface between the converters and the digital signal processing unit (DSP/FPGA). \\

 Previous works have exploited the sparsity of mmWave channels in the angle and delay
domains to design pilot-based channel estimation schemes \cite{You_2016,Schniter_2014}. Other works have considered pilot based channel and/or channel subspace  estimation in the context of hybrid MIMO mmWave systems with analog preprocessing \cite{Haghighatshoar_2016} and in the context of quantized MIMO mmWave systems with one-bit receivers \cite{Mo_2014}. Most of these works are based on a compressed sensing type of formulation and usually require a sufficiently high SNR per antenna before beamforming and/or a high degree of sparsity, which is not applicable for certain scenarios such as indoor and urban outdoor environments \cite{Rappaport_2013}. A maximum likelihood approach for blind and semi-blind estimation of massive MIMO mmWave channels has been presented in \cite{Neumann_2015} for Rayleigh fading channel models. Joint Bayesian channel-and-data estimation has been considered and analyzed in \cite{Wen_2015,Wen_Wu_2015,Steiner_2016} and shown to yield a large improvement  compared to training-based methods. However, this approach requires an iterative message passing algorithm applied to a sufficiently large system with significant complexity and generally strict assumptions on the prior distributions of the channel and data, and on the time and frequency synchronization, while convergence and optimality still  cannot be guaranteed. \\

 To address this issue, we present a  maximum likelihood approach for blind mmWave channel estimation that, unlike \cite{Neumann_2015}, takes into account the sparsity of these channels. Under this key property, we show that reliable estimation is possible at low SNR per antenna and unambiguous separation between users is still possible even though their channels are not orthogonal and prior distributions for the  channel and data are not available. An essential ingredient of our work compared to recent work is that the broadband array model includes the temporal shifts across the different elements, which is critical in the context of mmWave systems with large bandwidth and large array size. \\ 

Motivated by the advantages of one-bit receivers, in this paper we additionally aim at considering the blind channel estimation problem with one-bit observations. In contrast to the existing works \cite{Wen_2015,Steiner_2016,Mo_2016} that adopt the approximate message passing algorithm \cite{parker1} to solve this problem for a sufficiently large number of users, we aim here to develop a pure blind estimation approach based on non-convex optimization that works even for a moderate number of users. In fact, despite its theoretical potential \cite{Jianwen_20017}, the popular generalized approximate message passing algorithm \cite{rangan} and its extension to bivariate (data and channel) problems \cite{parker1} still require in practice a substantial  amount of users and pilots and a certain SNR level  to exhibit convergence and might suffer heavily from model mismatch. We show that the proposed  non-convex optimization approach with appropriate initialization for the blind estimation of weakly sparse channels can be applied for general cases (infinite resolution, 1-bit resolution, narrowband/broadband channel) while remaining robust to certain model assumptions, as for instance the type of modulation alphabet, as well as the time and frequency synchronization. \\

Our specific contributions consist of the following.
\begin{itemize}
\item We consider a general wideband channel model for mmWave massive MIMO where not only the phase shifts across the antennas are taken into account but also the temporal shifts. This last phenomenon, usually neglected in most related work, turns out to be essential in this context  and is due to the large bandwidth and the large distance between widely-spaced antennas in terms of wavelengths. 
\item We present a new maximum likelihood formulation for blind channel estimation based on $\ell_1$ regularization with ideal as well as with one-bit receivers. As the problem turns out to be non-convex and has to be solved iteratively, we  elaborate on the problem of finding a good initialization in closed form. The proposed formulation of the blind estimation problem does not require any time or frequency synchronization. 
\item  We derive iterative solutions for the maximum likelihood formulation which resemble the gradient
descent based thresholding algorithm. In the one-bit case, the iterative solution additionally makes use of the Expectation Maximization (EM) technique.  Through simulations, we demonstrate the benefits of the proposed approach, particularly compared to the common pilot-based method. On the other hand, we observe that the loss due to the 1-bit quantization is not significant in relevant scenarios.    
\item As a theoretical performance measure, we determine the Clairvoyant Cram\'er Rao lower bound (CRB) for both cases. Since for the one-bit case the  bound cannot be found in closed form, we  perform a Taylor series expansion of the CRB expression up to the second order in SNR.  
\end{itemize} 

The  next  section  introduces  the  channel  model  assumed  in 
the paper.  In Section~\ref{Subspace_Method}, the general blind channel estimation problem is formulated and approximately solved with a subspace approach. Then, taking this solution as an initialization, an iterative sparsity-based blind method is derived in Section~IV. In Section~V, the Clairvoyant Cram\'er Rao lower bound (CRB) is derived, providing a theoretical performance measure. These studies are also carried out for the one-bit case in Sections~VI and VII and a similar  iterative estimation algorithm is obtained by means of the EM algorithm. Finally, simulation results are given in Section~VIII to illustrate the performance gains compared with existing methods.  \\

\emph{Notation:}
Vectors and matrices are denoted by lower and upper case italic bold letters.  The operators $(\bullet)^\mathrm {T}$, $(\bullet)^\mathrm {H}$, $\textrm{tr}(\bullet)$ and $(\bullet)^*$ stand for transpose, Hermitian (conjugate transpose), trace, and complex conjugate, respectively.  The terms $\B{1}_M$ and $\B{I}_M$ represent the all ones vector and the identity matrix of size $M$, respectively. The vector $\boldsymbol{x}_i$ denotes the $i$-th column of a  matrix $\B{X}$ and $\left[\B{X}\right]_{i,j}$ denotes the ($i$th, $j$th) element, while $x_i$ is the $i$-th element of the vector $\B{x}$.  We represent the Hadamard (element-wise) and the Kronecker product of vectors and matrices by the operators "$\circ$" and "$\otimes$", respectively. Additionally, $\textrm{diag}(\boldsymbol{B})$ denotes a diagonal matrix containing only the diagonal elements of $\boldsymbol{B}$ and $\textrm{nondiag}(\boldsymbol{B})=\boldsymbol{B}-\textrm{diag}(\boldsymbol{B})$. Further, $\mathcal{F}\{\bullet\}$ and $\mathcal{F}^{-1}\{\bullet\}$ are the Fourier and inverse Fourier transform operators, respectively, used for both the continuous and discrete time domain depending on the context, and $\B{U}_{N}$ represents the normalized  DFT matrix of size $N$ with $\B{U}_{N}\B{U}_{N}^{\rm H}=\B{I}_N$. Finally, $\tilde{\B{x}}[n]$ is a time domain signal and $\B{x}[m]$ is the corresponding frequency domain signal.
\section{Wideband Channel Model}
In mmWave massive MIMO transmissions with $N$ antennas at the receiver, the wireless  propagation channel can be described by a sparse scattering  model, where  the $N$-dimensional channel vector $\B{h}_k$ of user $k$ consists of the superposition of $L_k \ll N$ multi-path components, typically including the line-of-sight (LOS) and some reflected paths \cite{Rappaport_2013,Akdeniz_2014}. The resulting ray-based channel model is variable in time and space and can be expressed as the following convolution   
\begin{equation}
\begin{aligned}
\tilde{\B{h}}_k(t)= \sum_{\ell=1}^{L_k} s_{\ell,k} \tilde{\B{a}}(\theta_{\ell,k},\varphi_{\ell,k},t) \ast \delta(t-t_{\ell,k}),
\end{aligned}
\end{equation}
where $s_{\ell,k}$ are the path coefficients (including path phase and strength), $\B{a}(\theta_{\ell,k},\varphi_{\ell,k},n)$ the array impulse response for the angles-of-arrival (AoA)  $\theta$ and $\varphi$ in a spherical coordinate system as a function of time, and $t_{\ell,k}$ the associated time delay. In the frequency domain, we have the channel transfer function
\begin{equation}
\begin{aligned}
  \B{h}_k(\omega)= \mathcal{F}\{\tilde{\B{h}}_k(t) \}= \sum_{\ell=1}^{L_k} s_{\ell,k} \B{a}(\theta_{\ell,k},\varphi_{\ell,k},\omega) {\rm e}^{-{\rm j}\omega {t_{\ell,k}} }.
\end{aligned}
\end{equation}
Sampling the delay-angle space on a grid, we introduce the angular array response matrix or dictionary $\B{A}(\omega)$, obtained by evaluating $\B{a}(\theta_{\ell,k},\varphi_{\ell,k},\omega)$ at a sufficiently
dense grid of angles (at least $N$) and, assuming the delays are integer multiple of the symbol period $1/B$, we can write the channel transfer function as\footnote{Later, in the simulations, we generate the delays and angles as continuous random variables not necessarily drawn from the constructed discrete  grid.}
\begin{equation}
\begin{aligned}
  \B{h}_k(\omega)= \big(\underbrace{[1,{\rm e}^{-{\rm j}{ \frac{\omega}{B} }}, \cdots,{\rm e}^{-{\rm j}{\frac{T_D\omega}{B} }}]  \otimes \B{A}(\omega)}_{\B{F}(\omega)} \big) \cdot \B{s}_k,
\end{aligned}
\end{equation}
where $\B{s}_k$ is a sparse vector containing the corresponding delay and AoA coefficients and $T_D$ is the maximum delay spread, i.e. $t_{\ell,k} \in\{0,1/B,\cdots,T_D/B \}$. \\ 

As an example, for a uniform planar array (UPA) of isotropic antennas with element spacing $d$ in wavelengths at the center frequency and dimensions $N_1 \times N_2$, we have the following broadband steering vector assuming all the frequencies propagate with the same speed and in the absence of antenna mutual coupling \cite{Brady_2015}:
\begin{equation}
\begin{aligned}
&\B{a}(\theta,\varphi,\omega)^{\rm T}\!=\!\frac{1}{\sqrt{N_1 N_2}}\!\left[1,\cdots\!, {\rm e}^{-{\rm j}{2 \pi}d \sin\theta(n_1 \sin\varphi+n_2 \cos\varphi)(1+\frac{\omega}{f_c})} , \right. \\
&\left. \cdots,  {\rm e}^{-{\rm j}{2 \pi}d \sin\theta({(N_1-1)}\sin\varphi+{(N_2-1)} \cos\varphi)(1+\frac{\omega}{f_c})} \right],
\end{aligned}
\end{equation}
where $f_c$ is the center frequency. The term $\frac{\omega}{f_c}$, which is often neglected in other work, accounts for the time shift across the antennas in the baseband and is essential in the context of mmWave massive MIMO with array dimensions of several wavelengths and a large available bandwidth.    However, if $\sqrt{{N_{i}}} \cdot\frac{B}{f_c} \ll 1$ for both $i=1$ and $i=2$, then the array response is nearly frequency independent and a possible dictionary $\B{A}$ for the UPA that covers the 3-dimensional space can be based on the normalized DFT matrices $\B{U}_{N_{1/2}}$ of size $N_{1}$ and $N_{2}$, providing a sufficient dictionary for the case of large uniform arrays:
 \begin{equation}
\begin{aligned}
\B{A}_{\rm UPA}=\B{U}_{N_{1}} \otimes \B{U}_{N_{2}}.
 \end{aligned}
\end{equation}
In such case, the matrices $\B{F}(\omega)$ fulfill the following property using the equality $(\B{D} \otimes \B{C}) \cdot (\B{B} \otimes \B{A})=(\B{D} \cdot \B{B}) \otimes (\B{C} \cdot \B{A}) $
\begin{equation}
\int_{-\pi B}^{\pi B} \B{F}(\omega)^{\rm H} \B{F}(\omega) {\rm d}\omega=   \B{I}_{T_D} \otimes  \B{A}^{\rm H} \B{A} =\B{I}_{T_D N}. 
\end{equation}
Apart from sparsity of the propagation channel, i.e., $L_k \ll N$, no assumption about the array geometry or the channel's statistical properties is actually needed for the derivation of the estimation method and the analysis. Later on, for the simulations, we assume for simplicity a uniform linear array (ULA) ($N_1=N,~N_2=1$) and that the AoAs $\theta_{\ell,k}$ are random and uniformly distributed between $0$ and $\pi$, while the multi-path coefficients $s_{\ell,k}$ are drawn from a complex Gaussian distribution with unit variance. \\

We assume a block fading channel with $K$ single antenna users  and $N$  receive antennas in the uplink. During a coherence time of $T$ symbols, each user constructs a block of $T$ data symbols using for instance single carrier pulse shaping or OFDM  processing and appends a cyclic prefix that is longer than the delay spread. After convolution with the channel, and discarding the cyclic prefix, the base station receives a block of $T$ sampled signal vectors. By means of the DFT of the $T-$length discrete-time signals, while indexing the discrete frequencies as
\begin{equation}
\omega_m=2\pi B \!\left(\!\frac{m}{T}-\left\lfloor m/T+1/2\right\rfloor\!\right)\!\in [-\pi B ,\pi B ), ~ 0 \leq m < T, 
\end{equation}
 the sampled received signal in the DFT domain reads as
\begin{equation}
    \B{y}[m] = \B{H}[m] \cdot \B{x}[m] + \B{z}[m]=\B{F}_m  \B{S}  \cdot  \B{x}[m]+ \B{z}[m],
\end{equation}
where $\B{z}[m] \in \mathbb{C}^N$ is the noise vector in the frequency domain having i.i.d. elements with unit variance,  $\B{S} =[\B{s}_1,\ldots,\B{s}_K] \in \mathbb{C}^{NT_D \times K}$, $\B{H}[m] =[\B{h}_1[m],\ldots,\B{h}_K[m]] \in \mathbb{C}^{N \times K}$ comprises the user channels  $\B{h}_k[m]$, $k=1, \ldots,K$, in the DFT domain that are assumed to be unknown and $\B{x}[0], \ldots, \B{x}[T-1]$ $\in \mathbb{C}^{K}$ is the transmitted data block in the DFT domain.
\section{Blind Channel Estimation: Approximative Subspace Method}
\label{Subspace_Method}
 Assuming the data from the users $x_{k}[m]$ are i.i.d. Gaussian distributed\footnote{From a practical point of view, the Gaussian assumption is not essential for the presented methods, as we obtain very similar simulation results with practical discrete input distributions as shown later in the simulation results.}  with variance $\rho$ (representing the SNR), then the conditional distribution of the received signal $\B{y}[m]$ given $\B{S}$ can be expressed as a multivariate Gaussian distribution with covariance matrices   $(\rho \B{H}[m]\B{H}[m]^{\rm H} + \B{I}) $:
\begin{equation}
\label{cond_ideal}
\begin{aligned}
     &p(\B{y}[0],\ldots,\B{y}[T-1]|\B{S}) = \\
     & \frac{{\rm exp}\left( -  \sum\limits_m \B{y}[m]^{\rm H} \left( \rho \B{F}_m \B{S} \B{S}^{\rm H}\B{F}_m^{\rm H} + \B{I} \right)^{-1} \B{y}[m] \right) }{\pi^{N\cdot T} \prod\limits_m\left|\rho \B{F}_m \B{S} \B{S}^{\rm H}\B{F}_m^{\rm H} + \B{I} \right|}.
\end{aligned}
\end{equation}
The corresponding log-likelihood function reads as
\begin{equation}
\begin{aligned}
    & L(\B{S}) = -\! \sum\limits_m \! \left(\! \! \B{y}[m]^{\rm H}  \! \left( \!\rho \B{F}_m \B{S}\B{S}^{\rm H} \B{F}_m^{\rm H} \!+\! \B{I} \!\right)^{-1} \!\!   \B{y}[m] \!\! \right) \!\\
   &~~~~~~~~-\! \sum\limits_m  \log\left|\rho \B{F}_m\B{S}\B{S}^{\rm H}\B{F}_m^{\rm H}\! +\! \B{I} \right|.
\end{aligned}
\end{equation}

Maximizing this log-likelihood function with respect to $\B{S}$ is a non-convex  problem, which in general cannot be solved in closed form. However, for the low SNR regime $\rho \ll 1$ or for the flat-fading channel case with $\B{F}_m =\B{F},\forall m$, a closed form solution can be obtained. In fact, for  $\rho \ll 1$, and using the first order Taylor approximations $\log\left|\rho \B{B}\! +\! \B{I} \right| \approx \rho{\rm tr}(\B{B})$ and $\left( \!\rho \B{B}\!+\! \B{I} \!\right)^{-1} \approx \B{I}-\rho \B{B}$, we obtain the approximation
\begin{equation}
\begin{aligned}
  &L(\B{S}) \approx \\
  & -\sum\limits_m \B{y}[m]^{\rm H} \B{y}[m] + \sum\limits_m \rho{\rm tr}\left(\B{S}^{\rm H}\B{F}_m^{\rm H}( \B{y}[m]^{\rm H} \B{y}[m]-\B{I})\B{F}_m\B{S} \right).
  \label{aprox_loglike}
\end{aligned}
\end{equation}
 One possible solution for the blind estimation of $\B{S}$ with rank $K$ that approximately maximizes the first order approximation of the log-likelihood function (\ref{aprox_loglike}) is given by 
\begin{equation}
    \hat{\B{S}} \!= \! {\rm argmax}_{\B{S}} \! \! ~~p(\B{y}[0],\ldots,\B{y}[T-1]|\B{S}) \!\approx\! \frac{1}{\sqrt{T\rho}}  \B{V}_{1:K}  \sqrt{[\B{\Sigma}_{1:K}]_+},
\label{sol_subspa}
\end{equation}
where $\B{V}_{1:K} $ are the $K$ eigenvectors corresponding to the $K$ largest eigenvalues $\B{\Sigma}_{1:K}$ of the matrix $\sum\limits_m \B{F}_m^{\rm H}(\B{y}[m]\B{y}[m]^{\rm H}- \B{I})\B{F}_m=\B{V}\B{\Sigma}\B{V}^{\rm H}$ and $[a]_+=\max(a,0)$. This approximate solution is actually the optimal solution for the narrowband case with $\B{F}_m =\B{F},\forall m$, being a unitary matrix \cite{Neumann_2015}. It should be also noticed that this solution is not unique and that multiplication from the right with any unitary matrix will also provide another valid solution.  The particular channel estimate in (\ref{sol_subspa}) is characterized by the fact that the users are assumed to be orthogonal to each other. Therefore the quality of the subspace-based estimate strongly depends on this  assumption, which requires a very large  number of antennas. In the following section, we provide a modification of the method exploiting the sparsity of the propagation scenario that can relax this assumption while performing well also for the frequency selective case. The method consists of an iterative algorithm for solving a non-convex problem, where this subspace solution can serve as an efficient initialization.     
\section{Blind Sparse Channel Estimation}

For the case of mmWave massive MIMO, with dozens or hundreds of antennas, the number of multi-path components from each user to the base station is usually much less than the number of antenna elements $N$. Therefore, assuming a large antenna array, the channel can be represented as
\begin{equation}
\B{H}[m]= \B{F}_m \cdot \B{S},
\end{equation}
where $\B{F}_m $  is the angular array dictionary at frequency index $m$ and $\B{S}$ is a sparse matrix representing the coefficients of the different multi-path components. Strictly    speaking, the matrix $\B{S}$ is not perfectly sparse since the path directions from the users to the base stations do not correspond exactly to the discrete  directions and delays defined by the dictionary $\B{F}_m$, resulting in a clustered type of sparsity known as the leakage phenomenon. Nevertheless, the assumption of sparsity becomes more valid the higher the number $N$ of base station antennas.

Based on these assumptions,  we can state the following $\ell_1$ regularized maximum likelihood problem for estimating the channel:
\begin{equation}
\begin{aligned}
    &\max\limits_{\B{S}} L(\B{S})- \lambda \left\| \B{S} \right\|_{1,1} = \\
   &~~~~ -\! \sum\limits_m \! \left(\! \! \B{y}[m]^{\rm H}  \! \left( \!\rho \B{F}_m \B{S}\B{S}^{\rm H} \B{F}_m^{\rm H} \!+\! \B{I} \!\right)^{-1} \!\!   \B{y}[m] \!\! \right) \!\\
   &~~~~-\! \sum\limits_m  \log\left|\rho \B{F}_m\B{S}\B{S}^{\rm H}\B{F}_m^{\rm H}\! +\! \B{I} \right| \!-\! \lambda \! \left\| \B{S} \right\|_{1,1} \! \\
   &\equiv 
    \label{opt_S} \max\limits_{\B{S}} \sum\limits_m \! \rho \B{y}[m]^{\rm H} \B{F}_m\B{S} \! \left( \!\rho \B{S}^{\rm H} \B{F}_m^{\rm H} \B{F}_m \B{S}\!+\! \B{I} \!\right)^{-1} \!\! \B{S}^{\rm H}  \B{F}_m^{\rm H} \B{y}[m] \!\!  \\
    &~~~~-  \sum\limits_m  \log\left|\rho \B{S}^{\rm H}\B{F}_m^{\rm H}\B{F}_m\B{S}\! +\! \B{I} \right| \!-\! \lambda \! \left\| \B{S} \right\|_{1,1}  , 
\end{aligned} 
\end{equation}
where the $\ell_{1,1}$ matrix norm $\left\| \B{S} \right\|_{1,1} = \sum_{n,k} |s_{n,k}|$ is used to encourage sparse solutions with regularization parameter $\lambda$, and the matrix inversion lemma has been applied in the last step. The advantage of this formulation is that apart from the channel sparsity, no further assumption is made  on the channel's distribution, which provides robustness  and allows for the estimation of even the channels of interfering users from adjacent base stations without any assumption about time and frequency synchronization. \\

Unfortunately, the problem in (\ref{opt_S}) is non-convex since the cost function is not concave. Nevertheless, it has been observed that in many cases, solving a non-convex problem locally with appropriate initialization using efficient gradient based methods \cite{Li_2016} can be very successful in practice. Therefore, we use the gradient descent based iterative thresholding algorithm to determine a local optimal solution, as derived in Appendix~A (see also \cite{Zymnis_2010,Hale_2008}):
\begin{equation}
\begin{aligned}
 &\B{S}_{\ell+1}= \\
 &{\rm exp} ( {\rm j} \angle (\B{S}_\ell -\mu \B{\Delta})) \circ \max \left( {\rm abs}( \B{S}_\ell -\mu \B{\Delta} ) - \mu \frac{\lambda}{2} \B{1} \cdot \B{1}^{\rm T}, \B{0} \right),
 \end{aligned}
 \label{fixed_iter}
\end{equation}
where the phase and the absolute value operations symbolized by $\angle(\bullet)$ and ${\rm abs}(\bullet)$, respectively, are applied element-wise to their matrix arguments and the gradient is given by  
\begin{equation}
\begin{aligned}
   \B{\Delta}  =& - \frac{\partial L(\B{S})}{\partial \B{S}^*} \\
   =& -\rho \sum_m \B{F}_m^{\rm H}\left( \rho \B{F}_m\B{S}\B{S}^{\rm H}\B{F}_m^{\rm H} + \B{I} \right)^{-1}  \B{y}[m]  \cdot \\
   &\B{y}[m]^{\rm H}  \left(\rho\B{F}_m\B{S}\B{S}^{\rm H}\B{F}_m^{\rm H} + \B{I} \right)^{-1} \B{F}_m \B{S} +  \\
    & \quad\quad\quad\quad\quad\quad  \rho \B{F}_m^{\rm H}\left(\rho\B{F}_m\B{S}\B{S}^{\rm H}\B{F}_m^{\rm H} + \B{I} \right)^{-1} \B{F}_m \B{S}. 
\end{aligned}
\label{gradient}
\end{equation}
As initialization for the iterative algorithm we take the subspace solution (\ref{sol_subspa}). The iterative method is summarized in Algorithm~\ref{Blind_alg}. We note that if one user happens to be inactive, then the corresponding column in the estimate of $\B{S}$ will be near zero, and thus the algorithm can be also used to determine the set of active users or find potential interferers. \\

 The solution of the optimization problem (\ref{opt_S}) is not unique, in the sense that any transformation of the form
\begin{equation}
  \B{H}'[m]=\B{H}[m]{\bf diag}([{\rm e}^{-{\rm j}2 \pi \frac{d_1}{T} m+{\rm j}\phi_1},\ldots,{\rm e}^{-{\rm j}2 \pi \frac{d_K}{T} m+{\rm j}\phi_K}])\B{\Pi},
  \label{perm}
\end{equation}
with any diagonal phase and time shift\footnote{Non-uniqueness of the solution in terms of time shift occurs only if the maximum delay spread $T_D$ is not reached.} matrix and any user permutation matrix  $\B{\Pi}$ provides an equally valid solution, a fact that reflects the non-uniqueness of assigning the channels to the user's indices. These ambiguities in terms of phase and time shifts and user assignment can be resolved easily by taking advantage of the finite-alphabet structure of the signals and information from the higher layers or by including a short training phase.
\begin{algorithm}[t]
\caption{Blind $\ell_1$ Regularized Channel Estimation}
\label{Blind_alg}
\begin{algorithmic}[1]
\STATE \textbf{Input:} $\B{y}[m]=\mathcal{F}\{\tilde{\B{y}}[n]\}/\sqrt{T}$, $0 \leq m \leq T-1$ in the DFT domain
\STATE \textbf{Initialize:} $ \B{V}\B{\Sigma}\B{V}^{\rm H} \leftarrow \sum\limits_m \B{F}_m^{\rm H}(\B{y}[m]\B{y}[m]^{\rm H}- \B{I})\B{F}_m$ \\
$\B{S}^{(0)} =  \frac{1}{\sqrt{T\rho}}  (\B{V}_{1:K}  \sqrt{[\B{\Sigma}_{1:K}]_+})$
\\     
$\mu > 0 $, $0<\beta <1$, $l\leftarrow 0$      
\REPEAT
\STATE  $i \leftarrow i+1$
\STATE Compute $\B{\Delta}^{(i-1)}$ from (\ref{gradient})
\STATE Gradient update:\\
 $\B{S}^{(i)} \leftarrow \B{S}^{(i-1)}-\mu \B{\Delta}^{(i-1)}$\\
\STATE Thresholding: \\
 $ \B{S}^{(i)} \! \leftarrow \!  {\rm exp}  ( {\rm j} \angle (\B{S}^{(i)}))\! \circ \! \max \left( {\rm abs}( \B{S}^{(i)}  ) \!-\! \mu \frac{\lambda}{2} \B{1} \cdot \B{1}^{\rm T}, \B{0} \right)$
\IF{$L(\B{S}^{(i-1)})- \lambda \left\| \B{S}^{(i-1)} \right\|_{1,1}  > L(\B{S}^{(i)})- \lambda \left\| \B{S}^{(i)} \right\|_{1,1} $} 
\STATE $\mu \leftarrow \beta\mu,\quad i\leftarrow i-1$\ENDIF
\UNTIL{desired accuracy for $\B{S}$ is achieved}
\STATE Back conversion:\\
   $\hat{\B{H}}[m]=\B{F}_m\cdot \B{S}$
\end{algorithmic}
\end{algorithm}
\subsection{Complexity Analysis}
Apart from determining the initialization, the most costly operation of   Algorithm~\ref{Blind_alg} in each iteration is the calculation of the gradient (\ref{gradient}), since the other steps consists mainly of element-wise operations with much lower complexity. The expression is composed of several repeated terms and therefore it is possible to reduce the number of arithmetic operations at the cost of some memory. Since the matrix to be inverted in (\ref{gradient}) contains a lower rank matrix, the matrix inversion lemma can be used to reduce the dimensionality and thus the complexity:
\begin{equation}
\left(\rho\B{F}_m\B{S}\B{S}^{\rm H}\B{F}_m^{\rm H} + \B{I} \right)^{\!-1} \B{F}_m \B{S} =  \B{F}_m \B{S} \left(\rho\B{S}^{\rm H}\B{F}_m^{\rm H}\B{F}_m\B{S} + \B{I} \right)^{\!-1}\!\!\!\!.
\end{equation}
Further, for each frequency index $m$, the gradient expression (\ref{gradient}) is comprised of the sum of two terms. The first term is a rank-one matrix obtained from the outer product of two vectors whereas the second summand $\rho \B{F}_m^{\rm H}\left(\rho\B{F}_m\B{S}\B{S}^{\rm H}\B{F}_m^{\rm H} + \B{I} \right)^{-1} \B{F}_m \B{S}$ is a full rank matrix  and requires therefore higher computational complexity.   
The complexity of computing this second term of the gradient expression (\ref{gradient})  is presented in Table~\ref{uniform}. \\

 The resulting total complexity per iteration is of the order $T\cdot o(4 N^2 T_D K +3NK^2 + K^3)$.  
It should be noted that the complexity analysis in Table~\ref{uniform} does not take into account the sparsity of $\B{S}$. This fact can be used to significantly reduce the complexity to the order of only $T\cdot(3NK^2)$, by neglecting the operations associated with the matrix multiplication $\B{F}_m\B{S}$. In addition, if the channel changes only slowly 
across consecutive blocks, an adaptive implementation of the algorithm as a stochastic gradient update with very few iterations per block can be envisaged, in order to further improve the performance while drastically lowering the number of iterations. Hence, we conclude that the complexity of the proposed algorithm is expected to be comparable to state-of-the-art pilot-based techniques. 
\begin {table}[thp]%
\caption {Complexity analysis per iteration and frequency Sample}
\label{uniform}\centering %
\begin{tabular}{|c|c|}
\hline %
Operation & Number of FLOPs \cite{hunger_2007} \\
\hline  %
\hline
$\B{P}_1 = \B{F}_m \B{S} \in \mathbb{C}^{N\times K}$  & $2N\cdot NT_D \cdot K-NK$ (worst case)   \\\hline %
$\B{P}_2 = \B{P}_1^{\rm H} \B{P}_1 \in \mathbb{C}^{K\times K}$  &   $NK^2+NK-K^2/2-K/2$  \\\hline %
$\B{P}_3 = (\rho \B{P}_2 +{\bf I})^{-1} \in \mathbb{C}^{K \times K}$ & $K^3+K^2+K$   \\\hline %
$\B{P}_4 = \B{P}_1 \B{P}_3 \in \mathbb{C}^{N\times K}$  & $2N \cdot K^2-NK$   \\\hline %
$\B{P}_5 = \B{F}_m^{\rm H} \B{P}_4 \in \mathbb{C}^{N T_D\times K}$  & $2N\cdot NT_D \cdot K-NT_D \cdot K$   \\\hline %
\hline
Total  & $o(4 N^2 T_D K +3NK^2 + K^3)$   \\\hline %
\end {tabular}
\end {table}
\section{Clairvoyant Cram\'er Rao lower bound (CRB)}
For a given channel $\B{H}[m]=[\cdots \B{h}_k[m]\cdots]= [\cdots \B{F}_m \B{s}_k\cdots]$, and assuming that the right singular vectors are perfectly known - to ensure the uniqueness of the spectral separation and thus the maximum likelihood solution - then the Fisher information matrix, leading to the so-called clairvoyant Cram\'er Rao lower bound with ``genie" side information, can be written as \cite{Kay_1993,cramer2004random}

\begin{equation}
\begin{aligned}
 &\!\!\!\!\!\!\!\!\left[\B{J}\right]_{T_DN(k-1)+i,T_DN(k'-1)+i'}  \\
 &=\sum\limits_{m=1}^T {\rm tr} \left(\B{Q}_m^{-1} \frac{\partial \B{Q}_m}{\partial s_{k,i}^*} \B{Q}_m^{-1} \frac{\partial \B{Q}_m} { \partial s_{k',i'} } \right) \\
 &= \sum\limits_m \rho^2 \cdot {\rm tr} \left(\B{Q}^{-1}_m \B{F}_m \B{s}_k {\bf e}_i^{\rm T} \B{F}^{\rm H}_m \B{Q}^{-1}_m \B{F}_m {\bf e}_{i'} \B{s}_{k'}^{\rm H} \B{F}^{\rm H}_m  \right)  \\
 &=\sum\limits_m \rho^2\cdot {\bf e}_i^{\rm T} \B{F}^{\rm H}_m \B{Q}^{-1}_m \B{F}_m {\bf e}_{i'}     \cdot \B{s}_{k'}^{\rm H} \B{F}^{\rm H}_m \B{Q}^{-1}_m \B{F}_m \B{s}_k,
\end{aligned}
\end{equation}
 with $\B{Q}_m=\rho \B{H}[m]\B{H}[m]^{\rm H}+\B{I}$. This can be written in a compact way as
 \begin{equation}
  \B{J}= \sum\limits_m \rho^2 \cdot \B{H}[m]^{\rm H} \B{Q}^{-1}_m \B{H}[m] \otimes \B{F}^{\rm H}_m \B{Q}^{-1}_m \B{F}_m .
     \label{fisher}
 \end{equation}
 At low SNR, i.e.,  $\B{Q}_m\approx \B{I}$, this can be approximated as 
 \begin{equation}
  \B{J}\stackrel{\rho \ll 1}{\approx}   \rho^2  \sum\limits_m \B{H}[m]^{\rm H}  \B{H}[m] \otimes (\B{F}^{\rm H}_m\B{F}_m).
\label{CRB_low_SNR}
 \end{equation}
 
Additionally, knowing the sparsity support of the channel, i.e.,  the indices of the non-zero elements of $\B{s}_k$, $\mathcal{S}=\{T_D N(k-1)+i| s_{k,i} \neq 0 \}$, a reduced-dimension Fisher information matrix can be obtained by taking the rows and the columns given by the subset $\mathcal{S}$ as follows
 \begin{equation}
   \tilde{\B{J}}=\B{J}_{\mathcal{S},\mathcal{S}}.
   \label{fisher2}
  \end{equation}
\section{Blind Estimation with one-bit observations}
In this section, we reconsider the blind estimation problem where the receiver only has access to the sign of the received signal in each dimension. The time domain received signal reads as 
\begin{equation}
\tilde{\B{r}}[n]=\frac{1}{\sqrt{2}}{\rm sign}({\rm Re} \{\tilde{\B{y}}[n]\})+\frac{\rm j}{\sqrt{2}}{\rm sign}({\rm Im} \{\tilde{\B{y}}[n]\}).
\end{equation}
We can thus write the conditional probability as follows \cite{mezghaniisit2008}:
\begin{equation}
\label{cond_onebit}
\begin{aligned}
&P(\tilde{\B{r}}[0],\ldots,\tilde{\B{r}}[T-1]|\B{S})=\\
&{\rm E}_{{\B{x}}[0],\ldots,\tilde{\B{x}}[T-1]}\left[P(\tilde{\B{r}}[0],\ldots,\tilde{\B{r}}[T-1]|\B{S},\tilde{\B{x}}[0],\ldots,\tilde{\B{x}}[T-1])\right]=\\
&\!\!\int\!\!\!\!\prod_{\mathcal{C}\in\{{\rm Re},{\rm Im} \}}\prod_{n,j}{\!\Phi\!\left(2 \mathcal{C}\{\tilde{r}_j[n]\} \mathcal{C}\{[\mathcal{F}^{-1}(\B{H}[m]\B{x}[m])]_j\} \right)} {\rm d}P(\tilde{\B{x}}[n]),
\end{aligned}  
\end{equation}
where $\Phi(z) = \frac{1}{\sqrt{2\pi}}\int_{-\infty}^{z}e^{-\frac{t^2}{2}}{\rm d}t=\frac{1}{\sqrt{2\pi}}\int_{0}^{\infty}e^{-\frac{(t-z)^2}{2}}{\rm d}t$ is the cumulative normal distribution function. Maximizing this conditional  probability with respect to $\B{S}$ is a difficult non-convex problem, besides the fact that the integrals cannot be obtained in closed form. Therefore, as done for the ideal case, we aim first at finding an approximate solution for the ML problem, which serves as an initialization for an iterative EM algorithm that exploits the sparsity of the multi-path coefficient matrix $\B{S}$. 
\subsection{Approximate Solution}
In a manner similar to the unquantized case, we derive an approximate solution of the maximum-likelihood problem
\begin{equation}
\begin{aligned}
 \max\limits_{\B{S}} P(\tilde{\B{r}}[0],\ldots,\tilde{\B{r}}[T-1]|\B{S}),
\end{aligned}
\end{equation}
which can serve as initialization for a more sophisticated iterative algorithm solving the non-convex problem and exploiting the sparsity. To this end, we use the results of Appendix~B to obtain the following first order approximation of the one-bit conditional probability:
\begin{equation}
\begin{aligned}
  &P(\tilde{\B{r}}[0],\ldots,\tilde{\B{r}}[T-1]|\B{S}) \approx \\
  & \frac{1}{4^{NT}}\! \left(\!\! 1\!  +  \!\frac{2\rho}{\pi} \!\sum\limits_m {\rm tr}\!\left(\B{S}^{\rm H}\B{F}_m^{\rm H}(\B{r}[m]\B{r}[m]^{\rm H}-\B{I})\B{F}_m\B{S} \right) \! \right),
  \end{aligned}
  \label{approx_onebit} 
\end{equation}
where 
 $\B{r}[m]=\mathcal{F}^{-1}\{\tilde{\B{r}}[n]\}$. We notice the similarity  to the unquantized case (\ref{aprox_loglike}). Again, a possible solution for the blind estimation of $\B{S}$ with rank $K$ based on the first order approximation of the log-likelihood function (\ref{approx_onebit}) is given by 
\begin{equation}
    \hat{\B{S}} \!=\!  {\rm argmax}_{\B{S}}   P(\B{r}[0],\ldots,\B{r}[T\!-\!1]|\B{S}) \!\approx\! \sqrt{\!\frac{\pi}{2T\rho}}  \B{V}_{\! 1:K}  \sqrt{[\B{\Sigma}_{1:K}]_+},
\label{sol_subspa_q}
\end{equation}
where $\B{V}_{1:K} $ are the $K$ eigenvectors corresponding to the $K$ largest eigenvalues $\B{\Sigma}_{1:K}$ of the matrix $\sum\limits_m \B{F}_m^{\rm H}(\B{r}[m]\B{r}[m]^{\rm H}-\B{I} )\B{F}_m=\B{V}\B{\Sigma}\B{V}^{\rm H}$.
\subsection{Blind EM Algorithm Exploiting Sparsity}
Now, we aim at deriving an iterative algorithm for broadband sparse channel estimation with one-bit receivers. Blind estimation with one-bit data is in general mathematically more challenging than the ideal case \cite{Shalom,mezghani_itg_2010}. Therefore,  we formulate the ML problem as an EM step similar to \cite{mezghani_itg_2010} with the hidden quantity $\B{y}[m]\B{y}[m]^{\rm H}$. The expectation step can be written as  (c.f. (\ref{opt_S})) 
\begin{equation}
\begin{aligned}
    &\max\limits_{\B{S}} L(\B{S}|\B{S}^{(i-1)})- \lambda \left\| \B{S} \right\|_{1,1}    \equiv \\
   & \label{opt_S1} \max\limits_{\B{S}} -\sum\limits_m \! {\rm tr} \Bigg[  {\rm E}\left[\B{y}[m]\B{y}[m]^{\rm H}|\B{r}[n],\B{S}^{(i-1)}\right]   \! \B{Q}_m^{-1}\!\! \Bigg] 
     \!\!    \\
    &~~~~~~~~~~~~~~~   \!\!  -  \sum\limits_m  \log\left| \B{Q}_m  \right| \!-\! \lambda \! \left\| \B{S} \right\|_{1,1}  ,
\end{aligned}   \
\end{equation}
where $\B{Q}_m=\rho \B{S}^{\rm H}\B{F}_m^{\rm H}\B{F}_m\B{S}\! +\! \B{I}$. Next, we derive an approximation for the expectation step (E-Step).

The optimal reconstruction of the unquantized  covariance matrix in (\ref{opt_S1}) from the quantized data is in general mathematically  intractable and hence  we have to resort to approximations. 
Let $\B{C}_y[n]$ and $\B{C}_r[n]$ be the temporal correlation matrices of $\tilde{\B{y}}[n]$ and $\tilde{\B{r}}[n]$, respectively. By the \emph{arcsine law} \cite{price,jacovitti1994estimation}, we have the following relationship between the quantized and the unquantized covariance matrices in the time domain 
\begin{equation}
 \B{C}_r[n] = \frac{2}{\pi} {\rm arcsin}\left({\rm diag}(\B{C}_y[0])^{\frac{1}{2}}\B{C}_y[n]{\rm diag}(\B{C}_y[0])^{\frac{1}{2}}\right),
\end{equation}
where the \emph{arcsin} function is applied element-wise and to the real part and imaginary part separately. In other words, the unquantized covariance matrix can be reconstructed from 1-bit observations up to an unknown diagonal scaling 
\begin{equation}
 \B{C}_y[n] =  {\rm diag}(\B{C}_y[0])^{\frac{1}{2}} {\rm sin}\left(\frac{\pi}{2} \B{C}_r[n]\right) {\rm diag}(\B{C}_y[0])^{\frac{1}{2}}.
\end{equation}

 Thus, we use the following approximation for the E-step, which is asymptotically unbiased and
consistent for a large number of observations:
\begin{equation}
\begin{aligned}
 &{\rm E}[\B{y}[m]\B{y}[m]^{\rm H}|\B{r}[n],\B{S}^{(i-1)}]  \\
 &\approx \! \frac{1}{\sqrt{T}} \! \left( \! \sum_{m=1}^{T} {\rm diag} \B{Q}_m^{(i-1)} \!\right)^{\!\!\frac{1}{2}} \!\! \mathcal{F}\!\left\{\! \sin\left( \frac{\pi}{2} \hat{\B{C}}_r[n] \right) \! \right\} \!\! \left( \! \sum_{m=1}^{T} {\rm diag} \B{Q}_m^{(i-1)} \!\right)^{\!\!\frac{1}{2}} \\
 &=\hat{\B{\Phi}}_y[m]^{(i-1)} ,
 \end{aligned} 
 \label{cov_y}
\end{equation}
with $\B{Q}_m^{(i-1)}=\rho \B{S}^{(i-1),\rm H}\B{F}_m^{\rm H}\B{F}_m\B{S}^{(i-1)}\! +\! \B{I}$ obtained from the last step and the sampled temporal covariance matrix\footnote{Window functions other than rectangular can be used to account for the limited length of the impulse response and hence improve performance.}
\begin{equation}
\hat{\B{C}}_r[n]=\left\{
\begin{array}{ll}
	  \frac{1}{\sqrt{T}}\mathcal{F}^{-1}\{\B{r}[m]\B{r}[m]^{\rm H}\} &  0 \leq n\leq T_D \\
	\B{0} & T_D <n< T -T_D \\
	\hat{\B{C}}_r[T-n]^{\rm H} & T-T_D \leq n \leq T-1.
\end{array}
\right.
\label{cov_r}
\end{equation}
Now, we recalculate the gradient of the log-likelihood function for the M-Step 
\begin{equation}
- \frac{\partial L(\B{S}|\B{S}^{(i-1)})}{\partial \B{S}^*}=  \B{\Delta}^{(i-1)},
\end{equation}
with
\begin{equation}
\begin{aligned}
   \B{\Delta}^{(i-1)}  = & -\rho \sum_m \B{F}_m^{\rm H} \B{Q}_m^{-1,(i-1)}   \hat{\B{\Phi}}_y[m]^{(i-1)}     \cdot \\
   &   \B{Q}_m^{-1,(i-1)} \B{F}_m \B{S}^{(i-1)} +  \rho  \B{F}_m^{\rm H}\B{Q}_m^{-1,(i-1)} \B{F}_m \B{S}^{(i-1)},
\end{aligned}
\label{gradient1}
\end{equation}
where $\hat{\B{\Phi}}_y[m]^{(i-1)}$ comes from (\ref{cov_y}). The resulting iterative method for blind channel estimation with one-bit observations is summarized in Algorithm~\ref{Blind_alg_q}. \\
\begin{algorithm}[t]
\caption{1-Bit Blind $\ell_1$ Regularized Channel Estimation}
\label{Blind_alg_q}
\begin{algorithmic}[1]
\STATE \textbf{Input:} $\B{r}[m]=\mathcal{F}\{\tilde{\B{r}}[n]\}/\sqrt{T}$, $0 \leq m \leq T-1$ in the DFT domain
\STATE \textbf{Initialize:}  $\mu > 0 $, $0<\beta <1$, $l\leftarrow 0$ \\
Calculate $\B{C}_r[n] $ from (\ref{cov_r}) and $\mathcal{F}\!\left\{\! \sin\left( \frac{\pi}{2} \B{C}_r[n] \right) \! \right\}$\\
$ \B{V}\B{\Sigma}\B{V}^{\rm H} \leftarrow \sum\limits_m \B{F}_m^{\rm H}(\B{r}[m]\B{r}[m]^{\rm H}-\B{I})\B{F}_m$ \\
$\B{S}^{(0)} =  \sqrt{\frac{\pi}{2T\rho}}  (\B{V}_{1:K}  \sqrt{[\B{\Sigma}_{1:K}]_+})$      
\REPEAT
\STATE  $i \leftarrow i+1$ \\
\STATE $\B{Q}_m^{(i-1)}=\rho \B{S}^{(i-1),{\rm H}}\B{F}_m^{\rm H}\B{F}_m\B{S}^{(i-1)}\! +\! \B{I}$
\STATE \underline{E-Step:} compute $\hat{\B{\Phi}}_y[m]^{(i-1)}$ from (\ref{cov_y})
\STATE \underline{M-Step:} Compute $\B{\Delta}^{(i-1)}$ from (\ref{gradient1}) 
\STATE Gradient update:\\
 $\B{S}^{(i)} \leftarrow \B{S}^{(i-1)}-\mu \B{\Delta}^{(i-1)}$\\
\STATE Thresholding: \\
 $ \B{S}^{(i)} \! \leftarrow \!  {\rm exp}  ( {\rm j} \angle (\B{S}^{(i)}))\! \circ \! \max \left( {\rm abs}( \B{S}^{(i)}  ) \!-\! \mu \frac{\lambda}{2} \B{1} \cdot \B{1}^{\rm T}, \B{0} \right)$
\IF{$L(\B{S}^{(i-1)})- \lambda \left\| \B{S}^{(i-1)} \right\|_{1,1}  > L(\B{S}^{(i)})- \lambda \left\| \B{S}^{(i)} \right\|_{1,1} $} 
\STATE $\mu \leftarrow \beta\mu,\quad i\leftarrow i-1$\ENDIF
\UNTIL{desired accuracy for $\B{S}$ is achieved}
\STATE Back conversion:   $\hat{\B{H}}[m]=\B{F}_m\cdot \B{S}$
\end{algorithmic}
\end{algorithm}
\section{Low SNR Clairvoyant Cram\'er Rao Bound (CRB) with One-bit Measurements}
We derive here the clairvoyant CRB for the one-bit case. As the conditional distribution is not available in closed form, we provide a low SNR approximation based on (\ref{approx_onebit}):

\begin{equation}
\begin{aligned}
 &\left[\B{J}^{\rm 1-bit}\right]_{T_DN(k-1)+i,T_DN(k'-1)+i'} \!=\!\!\!\!\!\!\sum\limits_{\tilde{\B{r}}[0,\cdots,T-1]}\!\!\!\!\!\!\!\!\! P(\tilde{\B{r}}[0,\cdots,T-1]|\B{S}) \\
 &   \frac{\partial \ln P(\tilde{\B{r}}[0,\cdots,T-1]|\B{S})}{\partial s_{k,i}^*} \frac{\partial \ln P(\tilde{\B{r}}[0,\cdots,T-1]|\B{S})} { \partial s_{k',i'} } \\
 &\approx \frac{1}{4^{NT}}\!\sum\limits_{\tilde{\B{r}}[0,\cdots,T-1]}\!\!\sum\limits_m (\frac{2\rho}{\pi})^2 \cdot {\bf e}_i^{\rm T} \B{F}^{\rm H}_m(\B{r}[m]\B{r}[m]^{\rm H}-\B{I})\B{F}_m \B{s}_k \cdot  \\
 & \quad\quad\quad\quad\quad\quad  \sum\limits_{m'}\B{s}_{k'}^{\rm H} \B{F}^{\rm H}_{m'}(\B{r}[m']\B{r}[m']^{\rm H}-\B{I})\B{F}_{m'}   {\bf e}_{i'}       
\end{aligned}
\end{equation}
Then, we concatenate all terms along the index $m$ in a compact way using the block diagonal matrices $\bar{\B{H}}_k={\bf diag}(\B{F}_0\B{s}_k,\cdots, \B{F}_{T-1}\B{s}_k)$ and $\bar{\bf E}_i={\bf diag}(\B{F}_0{\bf e}_i,\cdots, \B{F}_{T-1}{\bf e}_i)$, and the stacked vector $\bar{\B{r}}=[\B{r}^{\rm H}[0],\cdots, \B{r}^{\rm H}[T-1]]^{\rm H}$, to obtain

\begin{equation}
\begin{aligned} 
 &\left[\B{J}^{\rm 1-bit}\right]_{T_DN(k-1)+i,T_DN(k'-1)+i'} \!\approx \\
 &\frac{1}{4^{NT}}\! \sum\limits_{\bar{\B{r}}} \left(\frac{2\rho}{\pi}\right)^2  {\rm tr} (\bar{\bf E}_i^{\rm T} ( \bar{\B{r}}\bar{\B{r}}^{\rm H}-\B{I}) \bar{\B{H}}_k)      {\rm tr} (\bar{\B{H}}_{k'}^{\rm H} (\bar{\B{r}}\bar{\B{r}}^{\rm H}-\B{I})   \bar{\bf E}_{i'})     \\
 &=\frac{1}{4^{NT}}\!\sum\limits_{\bar{\B{r}}} \left(\frac{2\rho}{\pi}\right)^2    {\rm tr} (\bar{\bf E}_i^{\rm T} \bar{\B{U}}\bar{\B{U}}^{\rm H} ( \bar{\B{r}}\bar{\B{r}}^{\rm H}-\B{I}) \bar{\B{U}}\bar{\B{U}}^{\rm H} \bar{\B{H}}_k) \cdot  \\
 & \quad\quad\quad\quad\quad\quad  {\rm tr} (\bar{\B{H}}_{k'}^{\rm H} \bar{\B{U}}\bar{\B{U}}^{\rm H} (\bar{\B{r}}\bar{\B{r}}^{\rm H}-\B{I}) \bar{\B{U}}\bar{\B{U}}^{\rm H}  \bar{\bf E}_{i'})   \\
 &=\frac{1}{4^{NT}}\!\sum\limits_{\bar{\B{r}}} \left(\frac{2\rho}{\pi}\right)^2    {\rm tr} (\bar{\bf E}_i^{\rm T} \bar{\B{U}} ( \bar{\B{U}}^{\rm H}\bar{\B{r}}\bar{\B{r}}^{\rm H}\bar{\B{U}}-\B{I}) \bar{\B{U}}^{\rm H} \bar{\B{H}}_k) \cdot  \\
 & \quad\quad\quad\quad\quad\quad  {\rm tr} (\bar{\B{H}}_{k'}^{\rm H} \bar{\B{U}} (\bar{\B{U}}^{\rm H}\bar{\B{r}}\bar{\B{r}}^{\rm H}\bar{\B{U}}-\B{I}) \bar{\B{U}}^{\rm H}  \bar{\bf E}_{i'}) ,
\end{aligned}
\end{equation}
where $\bar{\B{U}}= \B{U}_T \otimes \B{I}_N$ and $\bar{\B{U}}^{\rm H}\bar{\B{r}} \in \{\pm \frac{1}{\sqrt{2}} \pm {\rm j}\frac{1}{\sqrt{2}}\}^{NT}$ is the time domain 1-bit signal.
 Now,  we make use of the following equality for any matrices $\B{B}$ and $\B{D}$:
\begin{equation}
\begin{aligned}
&\frac{1}{4^N} \sum_{\B{r} \in \{\pm \frac{1}{\sqrt{2}} \pm {\rm j}\frac{1}{\sqrt{2}}\}^N} \B{r}^{\rm H} \B{D} \B{r} \cdot \B{r}^{\rm H} \B{B} \B{r} = \\ 
 & \quad\quad\quad\quad\quad\quad\quad {\rm tr}( \B{D} \cdot {\rm nondiag}(\B{B})) + {\rm tr} (\B{D}) {\rm tr} (\B{B}),
\end{aligned}
\end{equation}
to finally obtain 
\begin{equation}
\begin{aligned} 
 &\left[\B{J}^{\rm 1-bit}\right]_{T_DN(k-1)+i,T_DN(k'-1)+i'} \!\approx \\
 &  \left(\frac{2\rho}{\pi}\right)^2  {\rm tr} (\bar{\B{U}}^{\rm H} \bar{\B{H}}_k\bar{\bf E}_i^{\rm T} \bar{\B{U}} {\rm nondiag}(  \bar{\B{U}}^{\rm H}\bar{\bf E}_{i'}\bar{\B{H}}_{k'}^{\rm H}\bar{\B{U}} ) ).    \\
\end{aligned}
\end{equation}

For the special case of flat fading ($\B{F}_m=\B{F}$, $\forall m$), we obtain 
\begin{equation}
\begin{aligned} 
 &\left[\B{J}^{\rm 1-bit}\right]_{T_DN(k-1)+i,T_DN(k'-1)+i'}^{\rm flat~channel} \!\approx \\
 &  T\left(\frac{2\rho}{\pi}\right)^2  {\rm tr} (\B{F} \B{s}_k{\bf e}_i^{\rm T} \B{F}^{\rm H} {\rm nondiag}(  \B{F}{\bf e}_{i'}\B{s}_{k'}^{\rm H}\B{F}^{\rm H} ) )    \\
 &= T\left(\frac{2\rho}{\pi}\right)^2 \left[ \B{s}_{k'}^{\rm H}\B{F}^{\rm H} \B{F} \B{s}_k{\bf e}_i^{\rm T} \B{F}^{\rm H} \B{F}{\bf e}_{i'}  - \right.  \\
  &\quad\quad\quad \left. {\rm tr} (\B{F} \B{s}_k{\bf e}_i^{\rm T} \B{F}^{\rm H} {\rm diag}(  \B{F}{\bf e}_{i'}\B{s}_{k'}^{\rm H}\B{F}^{\rm H} ) ) \right].
\end{aligned}
\label{J_1bit}
\end{equation}
After some matrix manipulations and defining $\B{h}_k=\B{F}\B{s}_k$, we have 
\begin{equation}
\begin{aligned}
 &\left[\B{J}^{\rm 1-bit}\right]_{T_DN(k-1)+i,T_DN(k'-1)+i'}^{\rm flat~channel} \!\approx \\
  & T\left(\frac{2\rho}{\pi}\right)^2 \Bigg(  \B{H}^{\rm H}\B{H}  \otimes \B{F}^{\rm H}\B{F}- \\
 & ({\bf I} \otimes \B{F}^{\rm H}) \left[ 
\begin{array}{ccc}
{\rm diag}(\B{h}_1\B{h}_1^{\rm H})& 	{\rm diag}(\B{h}_2\B{h}_1^{\rm H}) & \cdots\\
{\rm diag}(\B{h}_1\B{h}_2^{\rm H})& 	{\rm diag}(\B{h}_2\B{h}_2^{\rm H}) & \cdots\\
\vdots &  \vdots &  \ddots\\
\end{array}
 \right] ({\bf I} \otimes \B{F}) \Bigg).
\end{aligned}  
\end{equation}
We notice the similarity to the low SNR CRB (\ref{CRB_low_SNR}) in the ideal unquantized case to within a loss factor $4/\pi^2$ and the subtraction of the diagonal terms as the amplitude information cannot be used.
In the same way, assuming that the support of the sparse vectors $\B{s}_k$ is known, then the clairvoyant Fisher information matrix is obtained as in (\ref{fisher2}). 
\section{Simulation results with ideal and 1-bit observations}
\label{simulation}
As a performance measure for  evaluating the presented algorithms, we use the  correlation coefficient between
 the estimated and exact channel vector, given by
\begin{align}
\eta_{k} &=\frac{\max\limits_{-T_D\leq d_k\leq T_D}\left|\sum\limits_m \B{h}_k[m]^{\rm H} \hat{\B{h}}_k[m]  {\rm e}^{-{\rm j}2 \pi \frac{d_k}{T} m} \right|}{\sqrt{\sum\limits_m \left\|\B{h}_k[m]  \right\|_2^2 \sum\limits_m \left\|\hat{\B{h}}_k[m]  \right\|_2^2}}  \label{eta_k_exact} \\
& \leq \frac{\sum\limits_m|\B{h}_k[m]^{\rm H} \hat{\B{h}}_k[m] |}{\sqrt{\sum\limits_m \left\|\B{h}_k[m]  \right\|_2^2 \sum\limits_m \left\|\hat{\B{h}}_k[m]  \right\|_2^2}} \nonumber  \\
   &=  \frac{\sum\limits_m |\B{h}_k[m]^{\rm H} (\B{h}_k[m]+\B{e}_k[m]) |}{\sqrt{\sum\limits_m \left\|\B{h}_k[m]  \right\|_2^2\sum\limits_m\left\| \B{h}_k[m] + \B{e}_k[m]  \right\|_2^2}}  \nonumber\\
   &=\frac{|\B{h}_k[m]^{\rm H} (\B{h}_k[m]+\B{e}_k[m]) |}{\left\|\B{h}_k[m]  \right\|_2\left\|  \B{e}_k[m]  \right\|_2} \cdot \sqrt{\frac{\left\|\B{h}_k[m]  \right\|_2\left\|  \B{e}_k[m]  \right\|_2}{ \left\| \B{h}_k[m] + \B{e}_k[m]  \right\|_2^2}}  \cdot  \nonumber \\
   &~~~~~~~\cdot \sqrt{\frac{\left\|  \B{e}_k[m]  \right\|_2}{ \left\| \B{h}_k[m] \right\|_2}},
   \end{align}
where $\B{e}_k$  denotes the channel estimation error.  As a theoretical performance benchmark, we obtain the following expression based on the Fisher information matrix in (\ref{fisher}) and (\ref{fisher2}) or (\ref{J_1bit}) for the one-bit case: 
\begin{equation}
\begin{aligned}
 \eta_{{\rm CRB},k}  & \approx \frac{\sum\limits_m \left\|\B{h}_k[m]  \right\|_2}{\sqrt{\sum\limits_m \left\|\B{h}_k[m]  \right\|_2^2+\sum\limits_m \left\|\B{e}_k[m]  \right\|_2^2}}  \\
   & \approx \frac{1}{\sqrt{1+  \frac{\sum\limits_m {\rm tr}\left[\B{F}_m\B{\Pi}_k\tilde{\B{J}}^{-1}\B{\Pi}_k^{\rm H}\B{F}_m^{\rm H}\right]}{\sum\limits_m \left\|\B{h}_k[m]  \right\|_2^2}}} ,
\end{aligned} 
\label{CRB_eta}
\end{equation}
with the selection matrix for the corresponding user and its active coefficients 
\begin{equation}
\begin{aligned}
 \left[\B{\Pi}_k\right]_{i,j}=
 \left\{ \begin{array}{ll}
	1  & {\rm if}~ s_{k,i}\neq 0 ~ \wedge \\
     & j-i=\sum\limits_{k'=1}^{k-1}L_{k'}-\left|\left\{i'<i| s_{k,i'} = 0 \right\}\right| \\
	0  & {\rm otherwise,}
\end{array}
 \right.
\end{aligned}
\end{equation}
where we neglect in (\ref{CRB_eta}) the correlation factor between the estimation error $\B{e}_k[m]$ and the channel vector $\B{h}_k[m]$ for the unbiased estimator in the large system limit, i.e.,
\begin{equation}
\lim\limits_{N\rightarrow \infty} \frac{|\B{h}_k[m]^{\rm H} \B{e}_k[m] |}{\left\|\B{e}_k[m]  \right\|_2 \left\|\B{h}_k[m]  \right\|_2} \longrightarrow 0.
\end{equation}
Further, we approximate the squared norm of $\B{e}$ by the Cram\'er Rao Bound due to the law of large numbers.
\subsection{Narrowband Channel}
\label{simulation_narrow}
\begin{figure}[h]
\psfrag{x}[c][c]{$\eta$}
\psfrag{y}[c][c]{${\rm Pr}\left(\frac{|\hat{\B{h}}_k^{\rm H} \B{h}_k |}{\left\|\B{h}_k  \right\|_2\left\|\hat{\B{h}}_k  \right\|_2} \geq \eta \right)$}
\centerline{\epsfig{figure=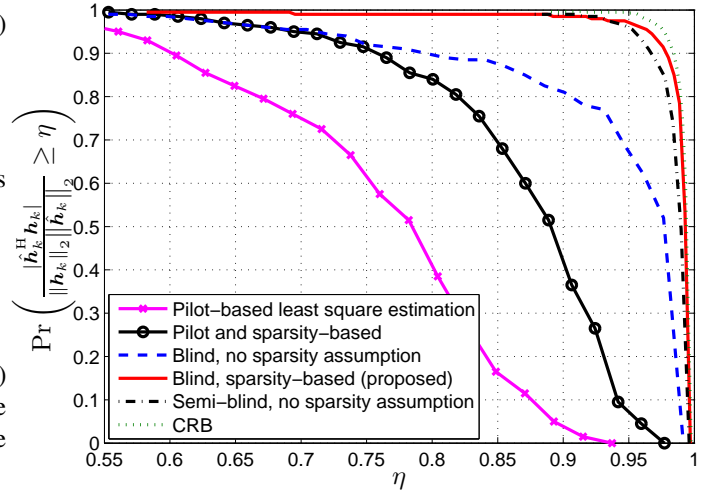,width=9cm}}
\caption{Narrowband channel estimation performance for $\rho=-12$dB, $N=32$, $K=2$, $L=3$, $T=1000$, $T_{\rm T}=10$.}
\label{fig_est}
\end{figure}

In the first simulation scenario, we assume a frequency flat scenario, i.e., $B \ll f_c$ and $T_D=0$, and we use $K=2$ users, a uniform linear array of size $N=32$ with half-wavelength element spacing, $L_k=3$ multi-path components per user and a coherence length of $T=1000$. The empirical complementary cumulative distribution of the correlation factor among all users and 100 channel realizations obtained by the proposed sparsity-based blind estimation, the subspace-based blind estimation (without any sparsity assumption) and the state-of-the-art pilot-based method are shown in Fig.~\ref{fig_est}. The $\ell_1$ regularization parameter is fixed at $\lambda=4$. We note that the generated  angles of arrival do not necessarily fall on the DFT grid, which means that the sparsity in the discretized angular domain does not perfectly hold. For the pilot-based methods, we use $K=2$ orthogonal pilot sequences $\B{X}_{\rm T} \in \mathbb{C}^{N\times T_{\rm T}}$ of length $T_{\rm T}=10$, and, when the sparsity is taken into account, the estimation is formulated as the popular regularized least squares problem for sparse reconstruction
\begin{equation}
\begin{aligned}
    &\min\limits_{\B{S}}  \left\| \B{FS}\B{X}_{\rm T}- \B{Y}_{\rm T} \right\|_2^2  + \lambda \! \left\| \B{S} \right\|_{1,1},
\end{aligned}    
\end{equation}
where $\B{Y}_{\rm T}$ is the received data during the pilot phase.  \\

 We first notice in Fig.~\ref{fig_est} the inferior performance of the conventional pilot-only based least squares channel estimation, even, surprisingly, if it takes sparsity into account. In fact, the blind methods clearly outperform pilot-only based methods, due to the low SNR per antenna ($\rho=-12$dB). Moreover, taking into account the sparsity of the propagation channel improves the blind estimation performance significantly and enables the algorithm to approach the clairvoyant Cram\'er Rao Bound. Concerning the non-uniqueness of the solution with respect to user permutations as described in (\ref{perm}), in this simulation we chose for both blind methods the permutation maximizing $\eta_1+\eta_2$.  

Furthermore, we compare the proposed approach with the semi-blind algorithm  presented in \cite{Neumann_2015}, where $T_{\rm T}$ symbols, denoted by $\B{X}_{\rm T}$, out of the $T$ sized block are dedicated for training. In this approach sparsity is not taken into account and the maximum likelihood optimization is formulated as (c.f. (\ref{opt_S}))
\begin{equation}
\begin{aligned}
    &\max\limits_{\B{H}} L(\B{H})-  \left\| \B{H}\B{X}_{\rm T}- \B{Y}_{\rm T} \right\|_2^2 = \\
   &\!\! -\! {\rm tr} \! \left(\! \! \B{Y}^{\rm H}_{\rm D } \! \left( \!\rho \B{H}\B{H}^{\rm H} \!+\! \B{I} \!\right)^{-1} \!\!   \B{Y}_{\rm D} \!\! \right) \!-\!(T-T_{\rm T}) \log\left|\rho \B{H}\B{H}^{\rm H}\! +\! \B{I} \right|  \\
   &\quad\quad\quad\quad\quad\quad\quad\quad\quad\quad\quad\quad\quad\quad\quad\quad- \left\| \B{H}\B{X}_{\rm T}- \B{Y}_{\rm T} \right\|_2^2,
\end{aligned}    
\end{equation}
where $\B{Y}_{\rm T}$ and $\B{Y}_{\rm D}$ represent the received signals corresponding to the known training block (commonly orthogonal sequences) and the unknown data block, respectively. For the simulation scenario, we take $K=2$ orthogonal pilot sequences of length $T_{\rm T}=10$ and solve the optimization problem using the gradient based method. Fig.~\ref{fig_est} shows that the proposed pure blind approach only exploiting the sparsity still outperforms the semi-blind approach \cite{Neumann_2015} using a pilot of length $T_{\rm T}=10$. This confirms the usefulness of the presented blind method in terms of reducing the training overhead. 

\begin{figure}[h]
\psfrag{x}[c][c]{$\eta$}
\psfrag{y}[c][c]{${\rm Pr}\left(\frac{|\hat{\B{h}}_k^{\rm H} \B{h}_k |}{\left\|\B{h}_k  \right\|_2\left\|\hat{\B{h}}_k  \right\|_2} \geq \eta \right)$}
\centerline{\epsfig{figure=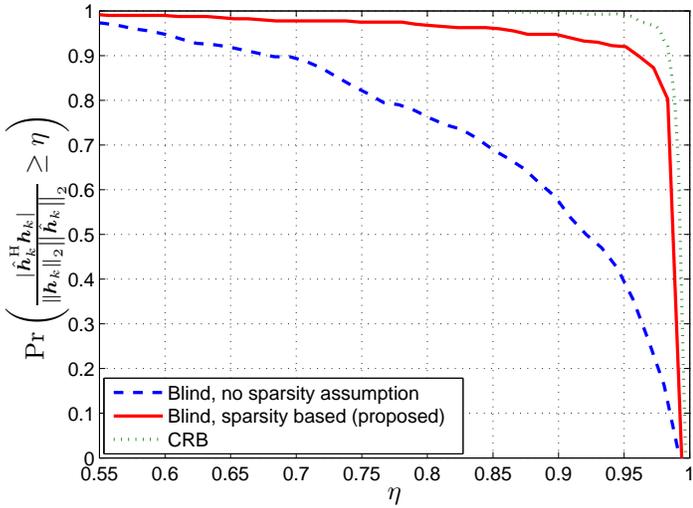,width=9cm}}
\caption{Narrowband channel estimation performance for $\rho=-12$dB, $N=64$, $K=4$, $L=3$, $T=1000$.}
\label{fig_est_64}
\end{figure}

In Fig.~\ref{fig_est_64}, we double the system size to $N=64$ antennas and $K=4$ users. The performance improvement of the proposed approach  compared to the pure subspace method is even  higher in this case. 
It is also expected that further performance advantages will be observed with higher numbers of users and larger antenna arrays.  

\begin{figure}[h]
\psfrag{x}[c][c]{$\eta$}
\psfrag{y}[c][c]{${\rm Pr}\left(\frac{|\hat{\B{h}}_k^{\rm H} \B{h}_k |}{\left\|\B{h}_k  \right\|_2\left\|\hat{\B{h}}_k  \right\|_2} \geq \eta \right)$}
\centerline{\epsfig{figure=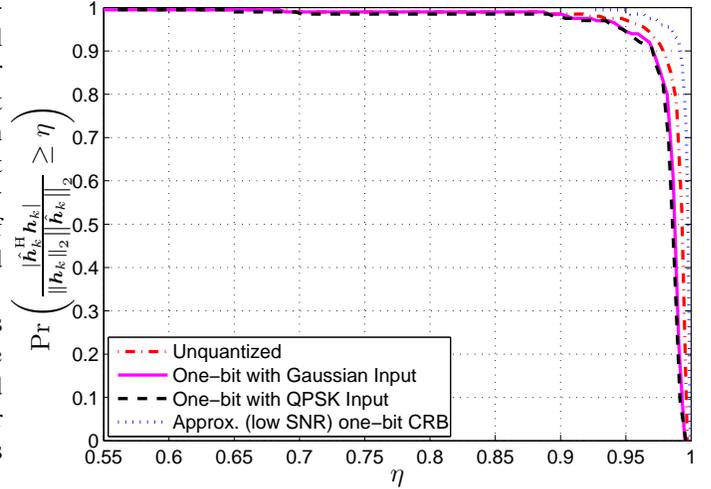,width=9cm}}
\caption{Estimation performance for $\rho=-12$dB, $N=32$, $K=2$, $L=3$, $T=1000$, $\lambda=4$ for the unquantized, and $\lambda=8$ for the one-bit case.}
\label{fig_est_q}
\end{figure}

In Fig.~\ref{fig_est_q} we evaluate the performance for the same scenario as the first with $N=32$ and $K=2$ while using one-bit quantizers and we observe that the performance is very close to the unquantized case, which confirms the advantage of using low resolution in the context of massive MIMO. In addition, the performance achieved using QPSK and Gaussian inputs is very similar, mainly due to the robustness of the Gaussian assumption used in the derivation of the algorithm, which maximizes the entropy given the variance.
\subsection{Wideband Channel}
\begin{figure}[h]
\psfrag{x}[c][c]{$\eta$}
\psfrag{y}[c][c]{${\rm Pr}\left(\eta_k \geq \eta \right)$}
\centerline{\epsfig{figure=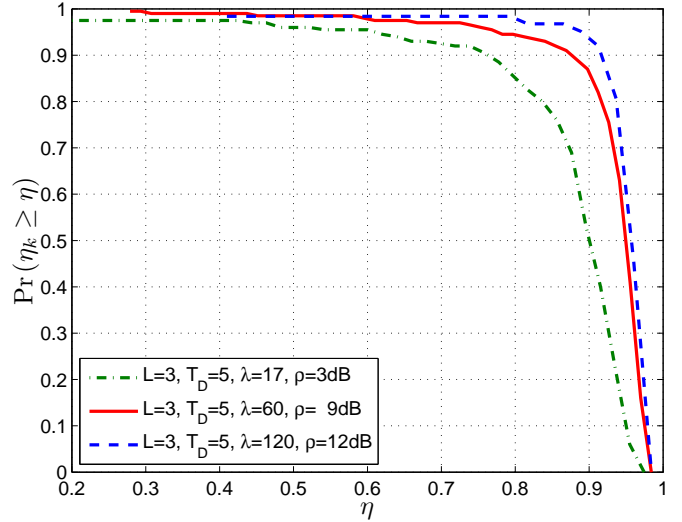,width=9cm}}
\caption{Wideband channel estimation performance for $\rho\in\{3,9,12\}$dB, $N=32$, $K=2$, $L=3$, $T=128$, $T_{D}=5$.}
\label{fig_est_sel}
\end{figure}

In the simulation that follows, we consider a wideband signal with bandwidth  $B=7$GHz and carrier frequency $f_c=60.5$GHz, corresponding to the unlicensed band at 57-64GHz. As in the previous setting,  we use $K=2$ users, a uniform linear array of size $N=32$ with half-wavelength element spacing, $L_k=3$ multi-path components, and a coherence length of $T=128$. The scattering parameters $s_{\ell,k}$ are chosen as independent circular-symmetric complex Gaussian random variables distributed as $s_{\ell,k} \sim \mathcal{CN}(0,1)$. The path delays $t_{\ell,k}$  are selected from a uniform distribution in the interval $[0,T_D]$, for $T_D=5$ symbol periods. The physical DoAs $\varphi_{\ell,k}$ for the antenna arrays are drawn from a uniform distribution in $[0,2\pi]$. Finally, the cumulative distribution of the correlation factor $\eta_k$ as defined in (\ref{eta_k_exact}) is obtained in Fig.~\ref{fig_est_sel} by simulating 100 independent channel realizations for $\rho\in\{3,9,12\}$dB.  
 
\begin{figure}[h]
\psfrag{x}[c][c]{$\eta$}
\psfrag{y}[c][c]{${\rm Pr}\left(\eta_k \geq \eta \right)$}
\centerline{\epsfig{figure=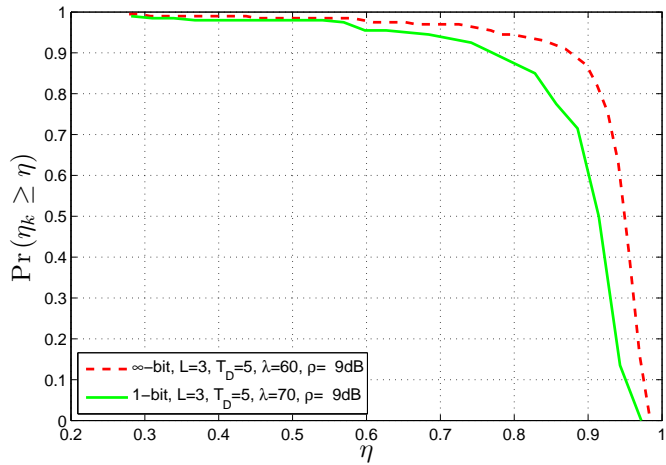,width=9cm}}
\caption{Wideband channel estimation performance with one-bit observations for $\rho=9$dB, $N=32$, $K=2$, $L=3$, $T=128$, $T_{D}=5$.}
\label{est_perf_sel_q}
\end{figure}

Next, in Fig.~\ref{est_perf_sel_q} we use the same wideband channels as in the previous case to determine the performance obtained by applying Algorithm~\ref{Blind_alg_q} to one-bit quantized observations. We note that the regularization parameter $\lambda$ is chosen to be higher for the one-bit case in order to account for the additional quantization effects. The obtained results are still encouraging for the one-bit case showing the potential of this simple solution for mmWave massive MIMO implementations. 

\section{Conclusion}
This work has considered the CRB and maximum likelihood approaches for blind massive MIMO channel estimation with ideal and one-bit receivers. An essential aspect of the assumed system model is that the array response is frequency dependent in the wideband regime due to the time shift across the antennas.  Based on the sparsity in the angular frequency domain, a non-convex $\ell_1$ regularized optimization problem is formulated and solved iteratively  with appropriately chosen initialization. The method significantly improves the spectral efficiency by dramatically reducing the overhead caused by pilot sequences and only exploits the sparsity property, and it is therefore robust to any type of statistical properties of the data and channels. Simulations
demonstrate that this maximum likelihood approach can
achieve a dramatic improvement in performance at low SNR. In fact, it allows blind separation of non-orthogonal channels just by taking advantage of the sparsity assumption. Finally, reliable channel estimation is shown to be still possible with simple one-bit receivers.

\section*{Appendix A}
First consider the subdifferentials of the $\ell_{1,1}$-norm $\left\| \B{S}\right\|_{1,1}$ with respect to any element $s_{i,j}$ using  Wirtinger's calculus: 
\begin{equation}
  \frac{\partial |z|}{\partial z}= \frac{\partial \sqrt{z\cdot z^*}}{\partial z}=\frac{z^*}{2\sqrt{z\cdot z^*}}=\frac{1}{2} {\rm e}^{-{\rm j} \angle(z)}.
\end{equation}
Therefore, we get the subdifferentials 
\begin{equation}
 \partial_{s_{i,j}} \left\| \B{S}\right\|_{1,1} \in \left\{
\begin{array}{cc}
	\frac{1}{2} {\rm e}^{-{\rm j} \angle(s_{i,j})} & \textrm{for~} s_{i,j} \neq 0, \\
	\{ \frac{1}{2} {\rm e}^{-{\rm j} \phi}  |\forall \phi \} & \textrm{for~} s_{i,j} = 0.
\end{array}\right.
\end{equation}
Then, the KKT conditions of the optimization problem (\ref{opt_S}) can be written as
\begin{equation}
 -\partial_{s_{i,j}} L(\B{S}) \in \left\{
\begin{array}{cc}
	-\frac{\lambda}{2} {\rm e}^{-{\rm j} \angle(s_{i,j})} & \textrm{for~} s_{i,j} \neq 0, \\
	\{ \frac{\lambda}{2} {\rm e}^{-{\rm j} \phi}  |\forall \phi \} & \textrm{for~} s_{i,j} = 0.
\end{array}\right.
\label{KKT}
\end{equation}
Considering now the following fixed point equation
\begin{equation}
\begin{aligned}
 &\B{S}= \\
 &{\rm exp} ( {\rm j} \angle (\B{S} -\mu \B{\Delta})) \circ \max \left( {\rm abs}( \B{S} -\mu \B{\Delta} ) - \mu \frac{\lambda}{2} \B{1} \cdot \B{1}^{\rm T}, \B{0} \right),
 \end{aligned}
 \label{fixed}
\end{equation}
with $\B{\Delta}^*=-\nabla_{\B{S}}L(\B{S})$ and $\mu>0$, any of its solutions is also a solution of the KKT conditions (\ref{KKT}). Solving (\ref{fixed}) using the fixed point iteration (\ref{fixed_iter}) with $\mu$ small enough converges provided that the gradient $\Delta$ is bounded and yields a local optimum for the optimization problem (\ref{opt_S}).
\section*{Appendix B}
For ease of illustration, we consider for simplicity the frequency flat case with $\B{y}=\B{H}\B{x}+\B{\eta}$ and $\B{r}=\frac{1}{\sqrt{2}}{\rm sign}({\rm Re} \{\B{y}\})+\frac{\rm j}{\sqrt{2}}{\rm sign}({\rm Im} \{\B{y}\})$. The frequency selective case can be tackled in a very similar way by constructing  the circulant convolution matrix based on the channel associated with the channel impulse response $\tilde{\B{H}}[n]$ and adopting again a matrix-vector notation. We first rewrite the conditional probability of the one-bit  output $\B{r}$ in (\ref{cond_onebit}) using the conditional probability of the unquantized output $\B{y}$ in (\ref{cond_ideal})
\begin{equation}
\begin{aligned}
&P(\B{r}|\B{H})=P({\rm Re}\{\B{r}\circ \B{y}\} \geq 0 \wedge {\rm Im}\{\B{r}\circ \B{y}\} \geq 0 |\B{H}) \\
&=\int_0^{\infty}\!\!\!\!\!\!\!\cdots\!\!\!\int_0^{\infty} p(\sqrt{2}\B{r} \circ \B{y}|\B{H}) {\rm d}\B{y}\\
&=\int_0^{\infty}\!\!\!\!\!\!\!\!\!\cdots \!\!\! \int_0^{\infty}\!\!\!\!  \frac{{\rm exp}(-2(\B{r} \circ \B{y} )^{\rm H}(\textbf{I}_N+ \rho\B{H}\B{H}^{\rm H})^{-1}(\B{r} \circ \B{y}))}{\pi^{N}\left|\textbf{I}_N+\rho\B{H} \B{H}^{\rm H}\right|}{\rm d}\B{y},
\label{cond_pro}
\end{aligned}
\end{equation}
where the integration is performed over the positive orthant of the complex hyperplane and $\boldsymbol{y}\circ \B{r}$ denotes a dimension-wise   vector product with ${\rm Re/Im}\{[\B{y}\circ \B{r}]_{i}\}={\rm Re/Im}\{y_{i}\}\cdot{\rm Re/Im}\{r_{i}\}$. Next, we compute the first order Taylor expansion of $P(\B{r}|\B{y})$ around $\rho=0$
\begin{equation}
\begin{aligned}
P(\B{r}|\B{H})\approx P(\B{r}|\B{H})_{\rho=0}+\rho P'(\B{r}|\B{H})_{\rho=0},
\end{aligned}
\end{equation}
where $P'(\B{y}|\B{x})_{\rho=0}$  is the first derivative of $P(\B{r}|\B{H})$ with respect to $\rho$. Clearly we have $P(\B{r}|\B{H})_{\rho=0}=1/4^N$ due to the i.i.d. noise. In order to calculate the first derivative, we use the first order approximation of the Gaussian conditional probability density\footnote{The following identities are useful: $\frac{\partial }{ \partial \rho} {\rm det} \B{Q}= {\rm det} \B{Q} \cdot {\rm tr}(\B{Q}^{-1} \frac{\partial \B{Q}}{ \partial \rho}) $ and $\frac{\partial  \B{Q}^{-1}}{ \partial \rho}=-  \B{Q}^{-1}\frac{\partial \B{Q}}{ \partial \rho} \B{Q}^{-1} $.} 
\begin{equation}
\begin{aligned}
p(\sqrt{2}\B{r} \circ \B{y}|\B{H}) \approx \frac{\exp(-\| \B{y}\|^2)}{\pi^N} \left(1-\rho {\rm tr}(\B{H} \B{H}^{\rm H})\right. \\
 \left. + 2 \rho {\rm tr} \{  (\boldsymbol{r} \circ \B{y})(\boldsymbol{r} \circ \B{y})^{\rm H}  \B{H} \B{H}^{\rm H}  \}\right).
 \end{aligned}
 \label{unq_cond_prob_app}
\end{equation}
Afterwards, we can show that
\begin{equation}
\begin{aligned}
\int_0^{\infty}\!\!\!\!\!\!\!\cdots\!\!\!\int_0^{\infty}  2\frac{{\rm e}^{-\| \B{y}\|^2}}{\pi^N} (r_i \circ y_i)(r_j \circ y_j)^{\rm *} {\rm d}\B{y}=\left\{
\begin{array}{ll}
\frac{1}{4^N}	 & \textrm{for } i=j\\
	2\frac{r_ir_j^*}{\pi4^N}	 & {\rm else.}
\end{array}
  \right. 
 \end{aligned}
\end{equation}
Thus,
\begin{equation}
\begin{aligned}
&\int_0^{\infty}\!\!\!\!\!\!\!\cdots\!\!\!\int_0^{\infty}  \frac{{\rm e}^{-\| \B{z}\|^2}}{\pi^N} {\rm tr}((\B{r} \circ \B{y})(\B{r} \circ \B{y})^{\rm H}   \B{H} \B{H}^{\rm H}) {\rm d}\B{y}=\\
& \frac{1}{4^N}  {\rm tr} \left( ( \textbf{I}_N + \frac{2}{\pi}{\rm nondiag}(\B{r}\B{r}^{\rm H}))\B{H} \B{H}^{\rm H} \right).
 \end{aligned}
 \label{int_approx}
\end{equation}
Combining (\ref{int_approx}), (\ref{unq_cond_prob_app}) and (\ref{cond_pro}), we get finally
\begin{equation}
\begin{aligned}
P'(\B{r}|\B{H})_{\rho=0}=\frac{1}{4^N}\frac{2}{\pi} {\rm tr} \left( \B{H}^{\rm H} \cdot \left(\B{r}\B{r}^{\rm H} - {\rm diag}(\B{r}\B{r}^{\rm H} )\right) \cdot \B{H}\right).
 \end{aligned}
\end{equation}
In conclusion, the first order approximation of $P(\B{r}|\B{H})$ is 
\begin{equation}
\begin{aligned}
P(\B{r}|\B{H})\approx\frac{1}{4^N}\left(1+\rho\frac{2}{\pi} {\rm tr} \left( \B{H}^{\rm H} \cdot \left(\B{r}\B{r}^{\rm H} - \B{I})\right) \cdot \B{H}\right)\right).
 \end{aligned}
 \label{fist_order_P(r)}
\end{equation}
This completes the proof for (\ref{approx_onebit}).  We notice also that since ${\rm tr}({\rm diag}(\B{B})\B{D})= {\rm tr}(\B{B}{\rm diag}(\B{D}))$ for any two matrices $\B{D}$ and $\B{B}$, we can express the approximation  (\ref{fist_order_P(r)}) also in a different way 
\begin{equation}
\begin{aligned}
P(\B{r}|\B{H})\approx\frac{1}{4^N}\left(1+\rho\frac{2}{\pi}   \B{r}^{\rm H} \cdot   {\rm nondiag}(\B{H}\B{H}^{\rm H} ) \cdot \B{r}\right).
 \end{aligned}
 \label{fist_order_P(r)_2}
\end{equation}



%
\bibliographystyle{IEEEtran}     
\bibliography{references}{}

\end{document}